\documentclass[doublecol]{epl2}
\usepackage{amssymb}
\usepackage{amsmath}

\newcommand{\eqb}{\begin{eqnarray}}
\newcommand{\eqe}{\end{eqnarray}}

\def\bes{\begin{subequations}}
\def\ees{\end{subequations}}

\newcommand{\cP}{\ensuremath{\mathcal{P}}}
\newcommand{\cT}{\ensuremath{\mathcal{T}}}

\title{$\cP\cT$-symmetric lattices with spatially extended gain/loss are generically unstable}
\shorttitle{$\cP\cT$-symmetric lattices with spatially extended gain/loss are generically unstable}

\author{D. E. Pelinovsky\inst{1} \and P.\ G.\ Kevrekidis \inst{2} \and D.\ J.\ Frantzeskakis \inst{3}}
\shortauthor{D. E. Pelinovsky \etal}

\institute{
\inst{1} Department of Mathematics, McMaster University, Hamilton, Ontario, Canada, L8S 4K1
\\
\inst{2} Department of of Mathematics and Statistics, University of
Massachusetts, Amherst, MA 01003-9305, USA
\\
\inst{3} Department of Physics, University of Athens, Panepistimiopolis,
Zografos, Athens 157 84, Greece
}

\abstract{
We illustrate, through a series of prototypical examples,
that linear parity-time ($\cP\cT$) symmetric lattices with extended
gain/loss profiles are generically unstable,
for any non-zero value of the gain/loss  coefficient. Our examples
include a parabolic real potential with a linear imaginary part
and the cases of no real and constant or linear imaginary potentials.
On the other hand, this instability can be avoided and the spectrum can be real
for localized or compact $\cP\cT$-symmetric potentials.
The linear lattices are analyzed through discrete Fourier transform techniques
complemented by numerical computations.
}

\begin{document}

\maketitle

{\it Introduction.} $\cP \cT$-symmetric quantum systems~\cite{R1}
(see also the review~\cite{R2}) have emerged as an intriguing
complex generalization of conventional quantum mechanics and have been a focus
for numerous investigations in theoretical physics and applied mathematics. The key premise is that fundamental
physical symmetries, such as parity $\cP$ and time reversal $\cT$, may be sufficient (in suitable
parametric regimes) to ensure that the eigenvalues of the Hamiltonian are real.
Thus, $\cP\cT$-symmetric Hamiltonians provide an alternative to the standard
postulate that the Hamiltonian operator must be Hermitian. For Hamiltonians associated with
a one-dimensional Schr\"{o}dinger operator with a complex potential $V(x)$, the constraint of
$\cP\cT$-symmetry requires that the potential satisfy $V(x) = \bar{V}(-x)$, i.e.,
$V(x)$ has a symmetric real part and an antisymmetric imaginary part.

Recently, the proposal that $\cP\cT$-symmetric systems can be physically
implemented in the framework of optics~\cite{dncprl}, was subsequently realized
in experiments utilizing active or passive $\cP\cT$-dimers~\cite{dncnat} and
periodic lattices~\cite{nat_rec}. Similar proposals for the
existence of a leaking dimer (in the presence of nonlinearity)
have been formulated in the atomic setting of Bose-Hubbard models \cite{R19}.
Further experimental investigations were concerned with electrical analogs of the
linear $\cP\cT$-symmetric system \cite{R21}.
Theoretical investigations have rapidly followed by examining such dimer-type
settings \cite{R22,R23,R25} and generalizations thereof, including ones where the gain/loss contributions appear
in a balanced form in front of the nonlinear term \cite{R31,R32,R33}.

Although dimers have been the principal workhorse on which the
investigation of an array of $\cP\cT$-symmetric systems
has been based, there are various works where
more elements were considered~\cite{R28,R28a}. $\cP\cT$-symmetric
solitons were studied in full nonlinear lattices
with a diatomic structure~\cite{Sukh,R30,KPZ}.
On the other hand, for the dimer monoatomic lattice, it is known that 
the $\cP\cT$-phase transition occurs at gain/loss
coefficient approaching zero when the number of lattice sites goes to infinity \cite{instability}.
In other words, infinite lattices consisting of dimers are linearly unstable
and discrete solitons cannot be robust in such lattices, contrary to their
counterparts in continuous models with periodic potentials \cite{R33,he}.

The purpose of the present work is to showcase the
dramatic differences between some prototypical discrete and continuum
$\cP\cT$-symmetric systems. In particular, we show that
linear $\cP\cT$-symmetric lattices with extended gain/loss are generically unstable:
{\it the eigenvalues of their associated Hamiltonians
are typically complex even if
they are known to be purely real for Hamiltonians associated with
the continuum analogue of the lattice}.
Our flagship example will be the standard quantum
harmonic oscillator incorporating a purely imaginary linear
potential~\cite{k22}. After discretization, the spectrum of
the discrete Schr\"{o}dinger operator includes infinitely many
complex eigenvalues, no matter how small the gain/loss coefficient is.

Other examples include the discrete Schr\"{o}dinger operator with no
real potential, and constant or linear imaginary potentials.
For these examples, we also show that either infinitely many isolated
eigenvalues or continuous spectral bands are complex inducing generic
instability of linear $\cP\cT$-symmetric lattices
in the presence of spatially extended gain/loss.

Nevertheless, we also briefly touch upon a setting where an infinite
lattice bearing $\cP\cT$-symmetry can have a real spectrum. This concerns
the case of {\it localized} or {\it compact} potentials. For example, a
$\cP\cT$-symmetric dimer embedded within the infinite lattice still enjoys
real spectrum for sufficiently small gain/loss coefficient~\cite{lepri}.
Therefore, localized or compact gain/loss may avoid the generic instability
scenario, which we study here for lattices with spatially extended gain/loss.

{\it A potential with a parabolic real part and a linear imaginary part.}
We start with the spectrum of the discrete Schr\"{o}dinger operator with
the potential $V_n = n^2 + i \gamma n$ (cf. Ref.~\cite{k22} for the continuous
version of this problem). Since $V_n = \bar{V}_{-n}$,
the potential is $\cP\cT$-symmetric. The relevant eigenvalue problem reads:
\begin{eqnarray}
E {u}_n= -\left(u_{n+1} + u_{n-1} -2 u_n\right) + \left(n^2+i \gamma n \right) u_n.
\label{eqn3}
\end{eqnarray}
Because the real part of the potential $V_n$ is bounded from below and confining as
$n \to \infty$, the spectrum of the linear lattice (\ref{eqn3}) is purely discrete
\cite{Zhang}. The continuous version of the linear lattice (\ref{eqn3})
takes the form:
$$
L = -\frac{d^2}{dx^2} + (x^2 + i \gamma x) = -\frac{d^2}{dx^2} +
\left(x + i \frac{\gamma}{2}\right)^2 + \frac{\gamma^2}{4}.
$$
Upon the change of variable $x \rightarrow x + i \gamma/2$, the eigenvalue problem
for $L$ is tantamount to the quantum harmonic oscillator that has a purely real spectrum
of eigenvalues located at $E = 2 m + 1 + \gamma^2/4$ for an integer $m \geq 0$ \cite{k22}.

\begin{figure}[tbp]
\begin{center}
\includegraphics[width=78mm,keepaspectratio]{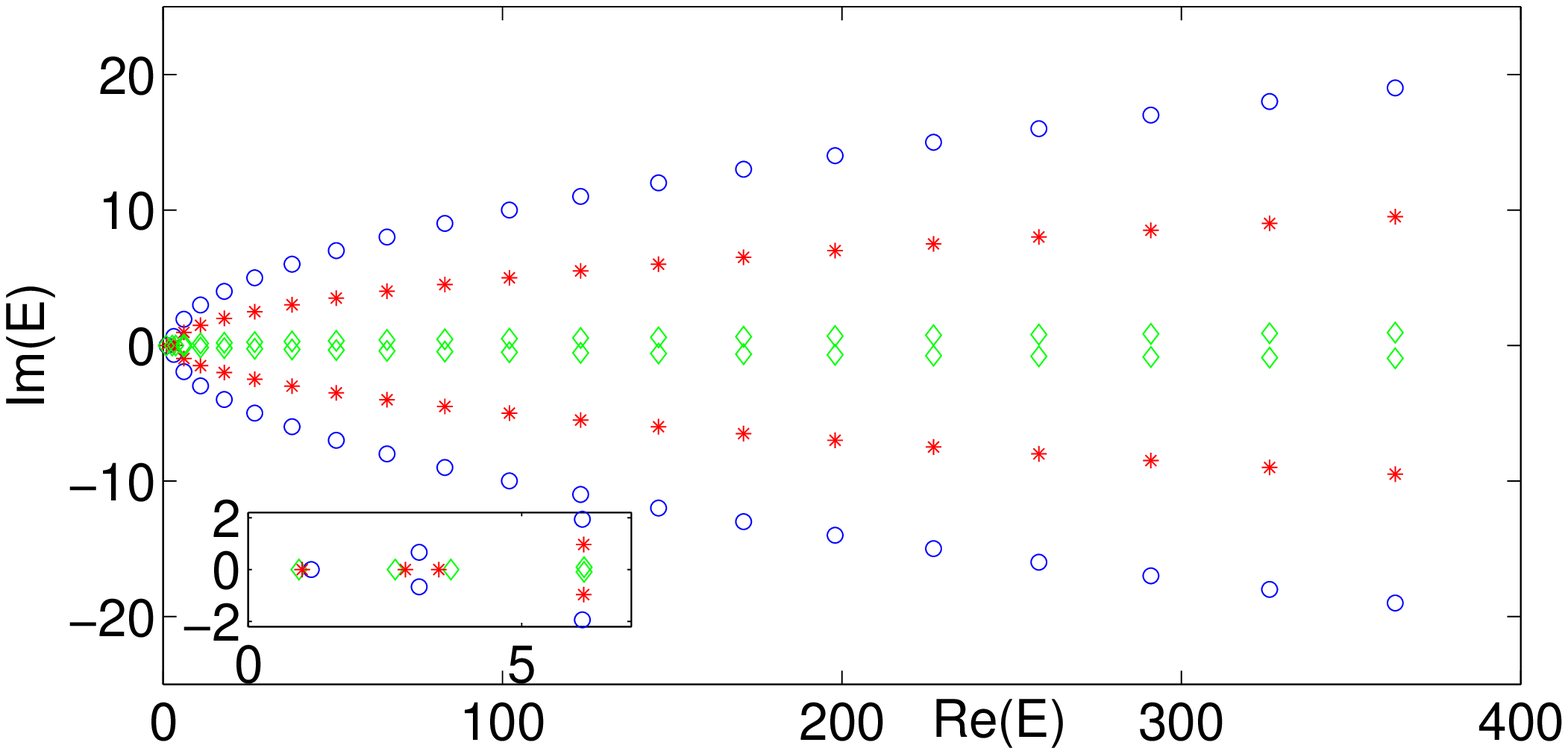}
\includegraphics[width=70mm,height=40mm]{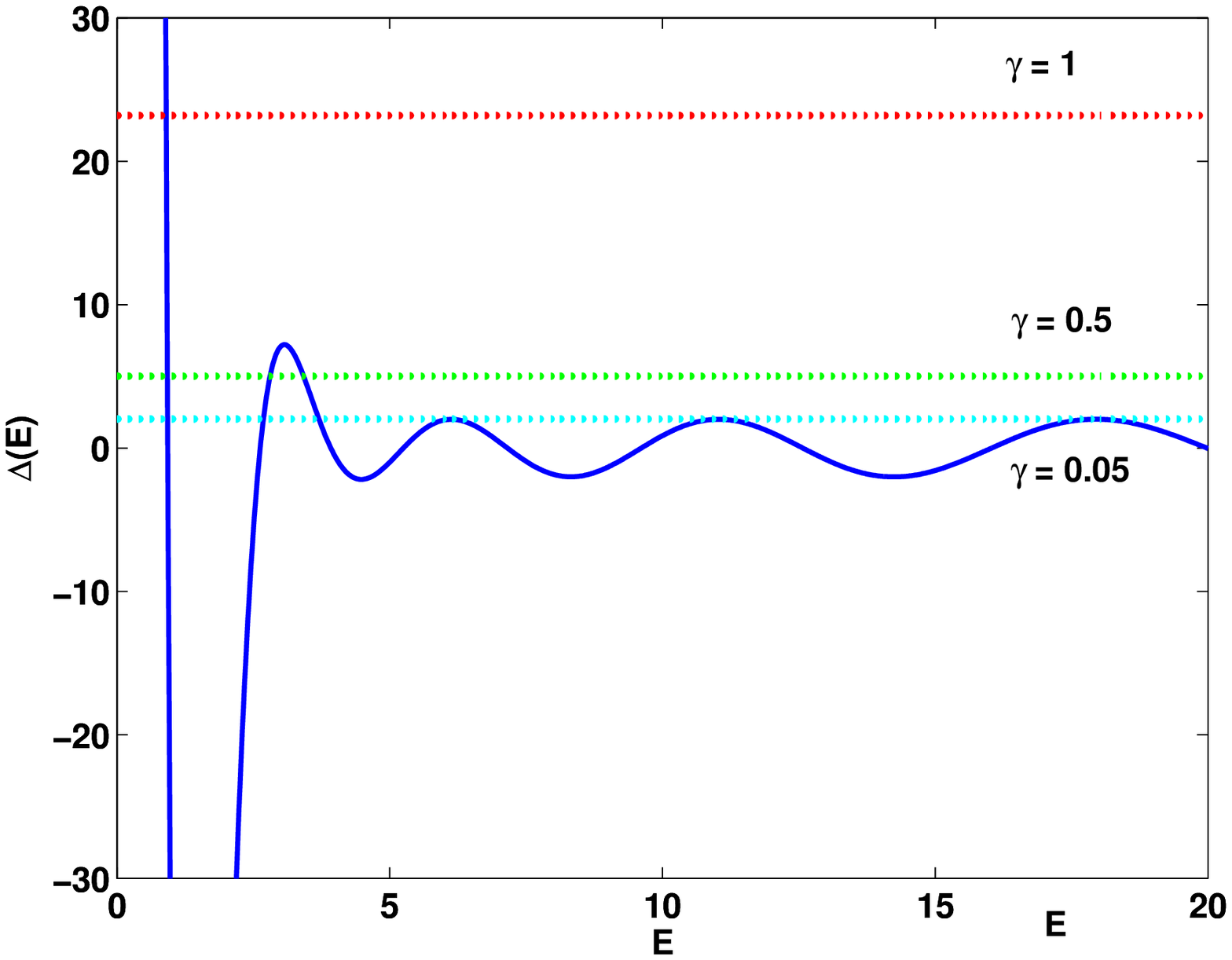}
\end{center}
\caption{(Color online) Top panel: Eigenvalues of the linear lattice (\ref{eqn3}), for the cases of $\gamma=1$
(blue circles), $\gamma=0.5$ (red stars) and $\gamma=0.05$
(green diamonds). The inset shows a blowup in the neighborhood of the
smallest eigenvalues, highlighting the existence of one real eigenvalue
for $\gamma=1$ and three real eigenvalues for the other two cases.
Bottom panel: trace of the monodromy matrix $\Delta(E)$ and
the levels of $2\cosh(\pi \gamma)$ for the three values of $\gamma$.}
\label{fig3}
\end{figure}

On the contrary, as shown in Fig.~\ref{fig3} (top), the discrete case is
significantly different in that the spectrum is {\it predominantly} complex.
The figure illustrates that, for different values of $\gamma$,
there are only a few eigenvalues on the real axis and an infinite number of complex eigenvalues.
The smaller the value of $\gamma$, the more eigenvalues are located on the real axis,
but it is always a finite number for any $\gamma \neq 0$. The spectrum of the
lattice (\ref{eqn3}) is of course real for the Hermitian case of $\gamma = 0$.

To confirm these numerical findings with analytic theory, we introduce the
discrete Fourier transform:
\begin{eqnarray}
u_n = \frac{1}{2\pi} \int_{-\pi}^{\pi} \hat{u}(k) e^{-i k n} dk.
\label{dFT}
\end{eqnarray}
Applying the discrete Fourier transform to the linear lattice (\ref{eqn3})
yields the differential equation in Fourier space:
\begin{eqnarray}
\frac{d^2 \hat{u}}{d k^2} + \gamma \frac{d \hat{u}}{d k}
+ \left[ E - 2 + 2 \cos(k) \right] \hat{u}(k) =0,
\label{eqn4}
\end{eqnarray}
where we are looking for $2\pi$-periodic functions $\hat{u}(k)$.
Applying the transformation $\hat{v}(k) = \hat{u}(k) e^{\gamma k/2}$,
we obtain the Mathieu equation:
\begin{eqnarray}
\frac{d^2 \hat{v}}{d k^2} + \left[E-2-\frac{\gamma^2}{4} + 2 \cos(k)\right]
\hat{v}=0.
\label{eqn5}
\end{eqnarray}
Now we have $\hat{v}(k+2 \pi) = e^{\pi \gamma} \hat{v}(k)$, that is,
we are looking for the Floquet multiplier $\mu_* = e^{\pi \gamma}$
of the monodromy matrix associated with the Mathieu equation (\ref{eqn5}).

As it is well-known \cite{Eastham}, the Floquet multiplier $\mu(E)$ is determined 
from the trace of the monodromy matrix $\Delta(E) = \mu(E) + \mu(E)^{-1}$. 
The function $\Delta$ diverges to positive infinity as $E \to -\infty$ and 
oscillates between values above $2$ and below $-2$ for $E > 0$. Moreover, 
the local maxima and minima of $\Delta(E)$ approach rapidly $\pm 2$
as $E \to \infty$, because the $\cos(k)$ potential in (\ref{eqn5}) is smooth.

The trace of monodromy matrix $\Delta(E)$ is shown in Fig.~\ref{fig3} (bottom) together
with constant levels of $2 \cosh(\pi \gamma)$ for the three values of $\gamma$.
For any given $\gamma > 0$, there are finitely many real roots $E$ of equation $\mu(E) = \mu_*$
and the number of real roots grows as $\gamma \to 0$ (it always includes at least one real root).
For instance, we have one root for $\gamma = 1$ and three roots for $\gamma = 0.5$ and $\gamma = 0.05$.
All other roots (infinitely many) are complex-valued. These roots bifurcate to complex numbers
via saddle-node bifurcations when $\mu_*$ is increased from $1$ (corresponding to $\gamma = 0$)
as $\gamma$ is increased.

{\it A potential with no real part and a linear imaginary part.} Let us now
drop the parabolic real potential and consider the linear eigenvalue problem:
\begin{eqnarray}
E u_n = -\left(u_{n+1} + u_{n-1} -2 u_n\right) + i \gamma n u_n.
\label{eqn1}
\end{eqnarray}
Applying the discrete Fourier transform (\ref{dFT}) yields
the first-order differential equation in Fourier space:
\begin{eqnarray}
\gamma \frac{d \hat{u}}{d k} + \left[ E - 2 + 2 \cos(k) \right]
\hat{u} =0.
\label{eqn2}
\end{eqnarray}
This equation can be exactly solved:
\begin{eqnarray}
\hat{u}(k) = \hat{u}(0) e^{\gamma^{-1} \left[ (2-E) k - 2 \sin(k)\right]}.
\end{eqnarray}
The $2\pi$-periodicity of the discrete Fourier transform $\hat{u}(k)$
gives now the eigenvalues $E = 2 + i \gamma m$, where $m$ is an arbitrary integer.
Hence, all the eigenvalues have an equidistant structure along the line
${\rm Re}(E)= 2$ and again there are infinitely many complex eigenvalues.

This conclusion can be checked directly from the difference equation (\ref{eqn1}).
If $E_0$ is an eigenvalue, then $E_0 + i \gamma m$ is also an eigenvalue for any
integer $m$ thanks to the discrete group of symmetry of the linear lattice
(\ref{eqn1}) with respect to translations in $n$.
Also $E_0 = 2$ is always an eigenvalue with the eigenvector $u_n = i^n I_n(2 \gamma^{-1})$,
where $I_n$ is the modified Bessel function. Note that the eigenvector
is decaying in $n$ because $I_n(2 \gamma^{-1}) \to 0$ as $n \to \pm \infty$
for fixed $\gamma > 0$.

\begin{figure}[tbp]
\includegraphics[width=78mm,keepaspectratio]{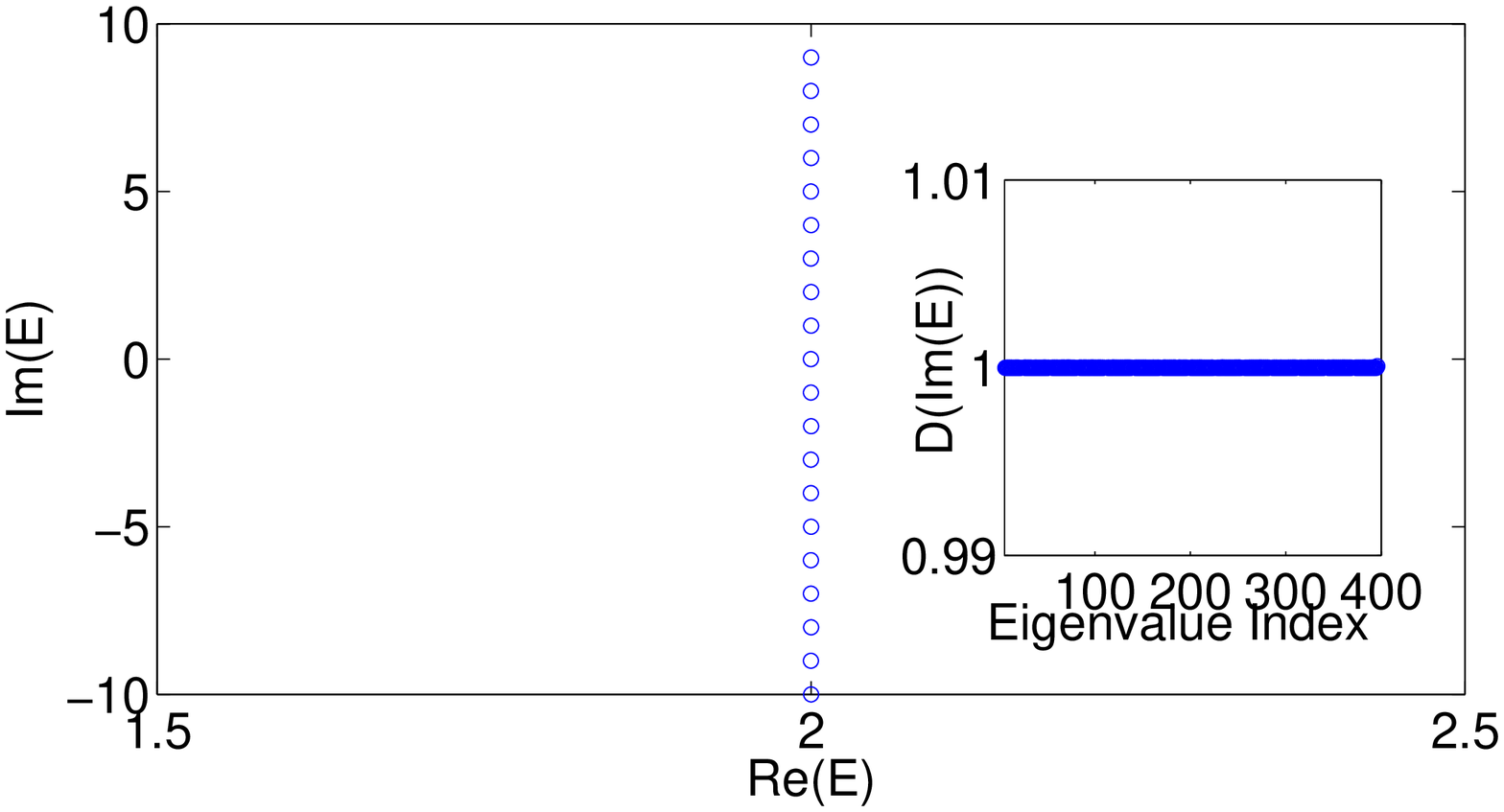}
\includegraphics[width=78mm,keepaspectratio]{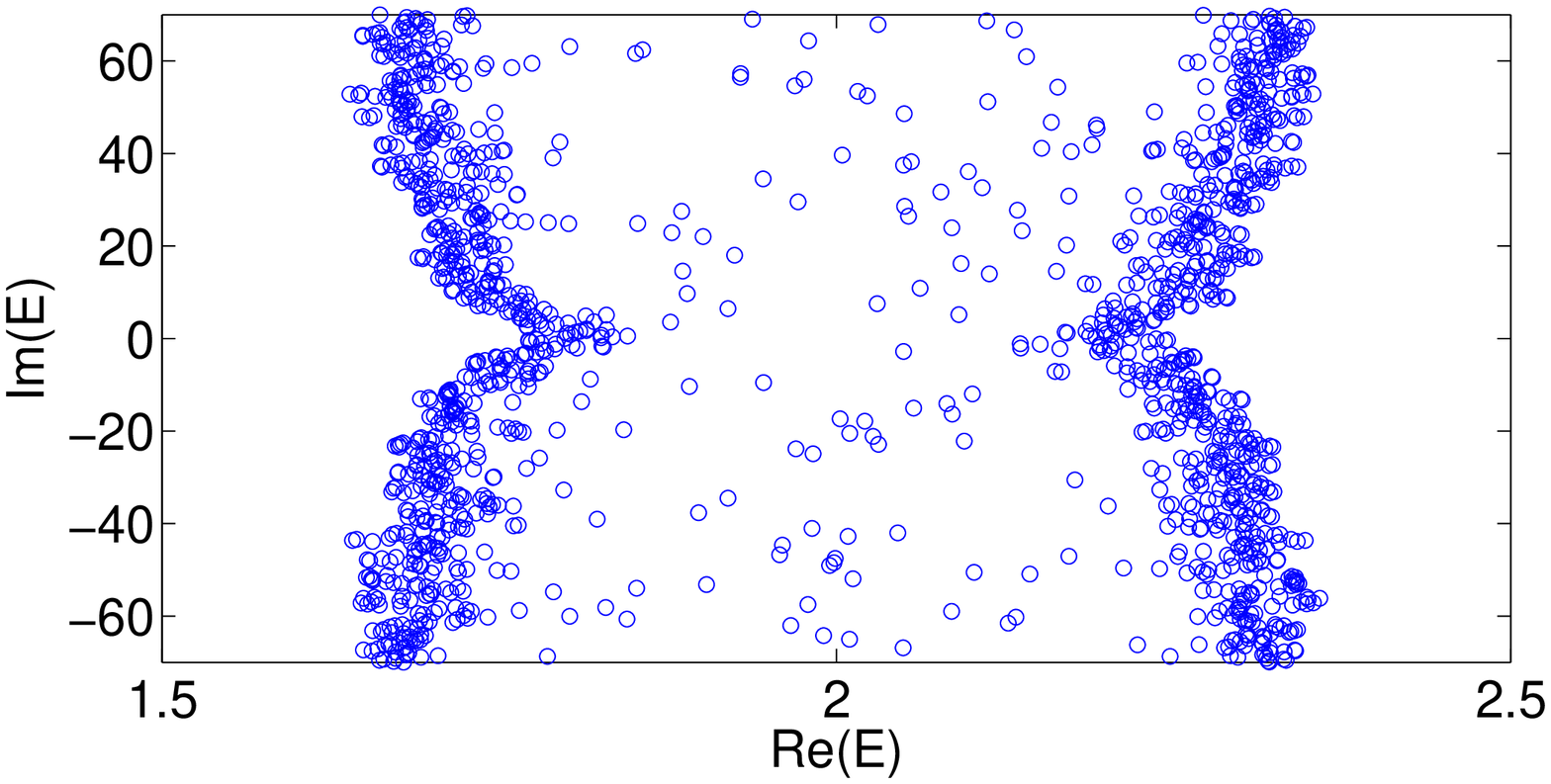}
\caption{(Color online) Eigenvalues of the linear lattice (\ref{eqn1}),
for the cases of $\gamma=1$ (top) and $\gamma=0.1$ (bottom); the inset
in the top panel shows the difference between
the imaginary parts of adjacent eigenvalues) }
\label{fig1}
\end{figure}

The sequence of equidistant eigenvalues along the line ${\rm Re}(E)= 2$
is illustrated in Fig.~\ref{fig1} for $\gamma = 1$ (top) and $\gamma = 0.1$ (bottom).
For $\gamma = 1$, we find that the equidistant eigenvalues $E = 2 + i \gamma m$
are the only eigenvalues of the linear lattice (\ref{eqn1}).
For smaller values of $\gamma$, we discover a new phenomenon, i.e.,
an appearance of additional parts of the spectrum of the
linear lattice (\ref{eqn1}). For $\gamma = 0.1$, the spectrum consists of
the continuous spectrum that fills space between two curves to the right and
left of the line ${\rm Re}(E) = 2$. For smaller values of $\gamma$,
the relevant spectral ``rectangle'' keeps expanding along the real parts
and contracting along the imaginary parts. The latter effect is induced by
the finite-size truncation effects, because the spectrum of the infinite
lattice always includes the equidistant eigenvalues $E = 2 + i \gamma m$.
The limit $\gamma \to 0$ is a singular limit of the differential equation (\ref{eqn2}),
when the only spectrum of the linear lattice (\ref{eqn1}) with $\gamma = 0$
is continuous and located on the real axis at $[0,4]$.

We now prove that the sequence of equidistant eigenvalues along the line ${\rm Re}(E)= 2$
is the only part of spectrum of the linear lattice (\ref{eqn1}) for sufficiently large values of
$\gamma$. Let $E = 2 + \gamma \lambda$ and rewrite the spectral problem (\ref{eqn1})
in the perturbed form:
\begin{eqnarray}
(\lambda - i n) u_n = - \gamma^{-1} \left(u_{n+1} + u_{n-1}\right).
\label{eqn6}
\end{eqnarray}
The limit $\gamma \to \infty$ corresponds to the limit of weak coupling of the lattice
and recovers the sequence of eigenvalues at $\lambda = i m$ for an integer $m$. Now,
let $\lambda$ be any complex number different from $\{ i m \}_{m \in \mathbb{Z}}$.
Rewriting (\ref{eqn6}) as $u = \gamma^{-1} K(\lambda) u$, we can see that the operator $K(\lambda)$
is bounded and therefore, there exists a sufficiently large value of $\gamma$ such that this $\lambda$
cannot be in the spectrum of the linear lattice (\ref{eqn6}). At the same time,
$\lambda = i m$ for any integer $m$ is a simple eigenvalue for $\gamma = \infty$ that
persists at $\lambda = im$ for any large $\gamma$.
This analytical argument shows that no continuous spectrum exists for sufficiently large
$\gamma$, so that there is a finite value of $\gamma=\gamma_0$, for which a bifurcation occurs in
the spectrum of the linear lattice (\ref{eqn1}). Mathematical analysis of this bifurcation
is beyond the scope of this work.

{\it A potential with no real part and a piecewise constant imaginary
part.} Let us now consider the piecewise constant potential in
the linear eigenvalue problem:
\begin{eqnarray}
E u_n = -\left(u_{n+1} + u_{n-1} -2 u_n\right) + i \gamma {\rm sign}(n) u_n.
\label{eqn7}
\end{eqnarray}
To solve this linear problem, we introduce $\theta(E)$ as a complex-valued root
of the dispersion relation
\begin{equation}
\label{disp-rel}
E = 2 - 2 \cos(\theta) + i \gamma.
\end{equation}
Since $\cos(\theta)$ is $2\pi$-periodic and even,
the root of the dispersion relation (\ref{disp-rel})
is uniquely determined in the semi-opened half-strip ${\rm Re}(\theta) \in [-\pi,\pi)$
and ${\rm Im}(\theta) > 0$ for any $\gamma > 0$ and $E \in \mathbb{C}$
with ${\rm Im}(E) \neq \gamma$. Note that ${\rm Im}(\theta) = 0$ if ${\rm Im}(E) = \gamma$. 
For any ${\rm Im}(E) \neq \gamma$, the eigenstate of 
the linear lattice (\ref{eqn7}) are represented by the exponentially
decaying function
\begin{eqnarray}
u_n = \left\{ \begin{array}{c} e^{i \theta(E) n}, \quad n \geq 0, \\
e^{i \theta^*(E) n}, \quad n \leq 0, \end{array} \right.
\end{eqnarray}
where $\theta^*(E)$ is found from (\ref{disp-rel}) with the change
$\gamma \to -\gamma$ (and the same $E$).
The linear lattice (\ref{eqn7}) is satisfied for any $n \neq 0$, whereas
the equation at $n = 0$ gives another constraint on $\theta(E)$:
\begin{equation}
\label{disp-const}
e^{-i \theta(E)} - e^{-i \theta^*(E)} = i \gamma.
\end{equation}
If ${\rm Re}(E) \neq 2$, we find from (\ref{disp-rel}) and
(\ref{disp-const}) that ${\rm Im}(\theta(E)) = {\rm Im}(\bar{\theta}(E))$,
which is impossible because they have opposite signs, hence ${\rm Re}(E) = 2$.
Further studies of this system of equations with ${\rm Re}(E) = 2$
show that the only solution exists for ${\rm Im}(E) = 0$ and corresponds to
\begin{eqnarray}
\theta(E) = -\frac{\pi}{2} + i {\rm arcsin}\left(\frac{\gamma}{2}\right).
\end{eqnarray}
This construction gives the only eigenvalue of the linear lattice (\ref{eqn7}).
In addition, there are always two branches of the continuous spectrum
located at ${\rm Im}(E) = \pm \gamma$ for ${\rm Re}(E) \in [0,4]$,
which correspond to the oscillatory non-decaying eigenfunctions with $\theta(E) \in \mathbb{R}$.
We conclude that the only eigenvalue of the linear lattice (\ref{eqn7}) is stable,
but the branches of the continuous spectrum are nevertheless unstable for any
non-zero value of $\gamma$.

\begin{figure}[tbp]
\begin{center}
\includegraphics[width=70mm,height=50mm]{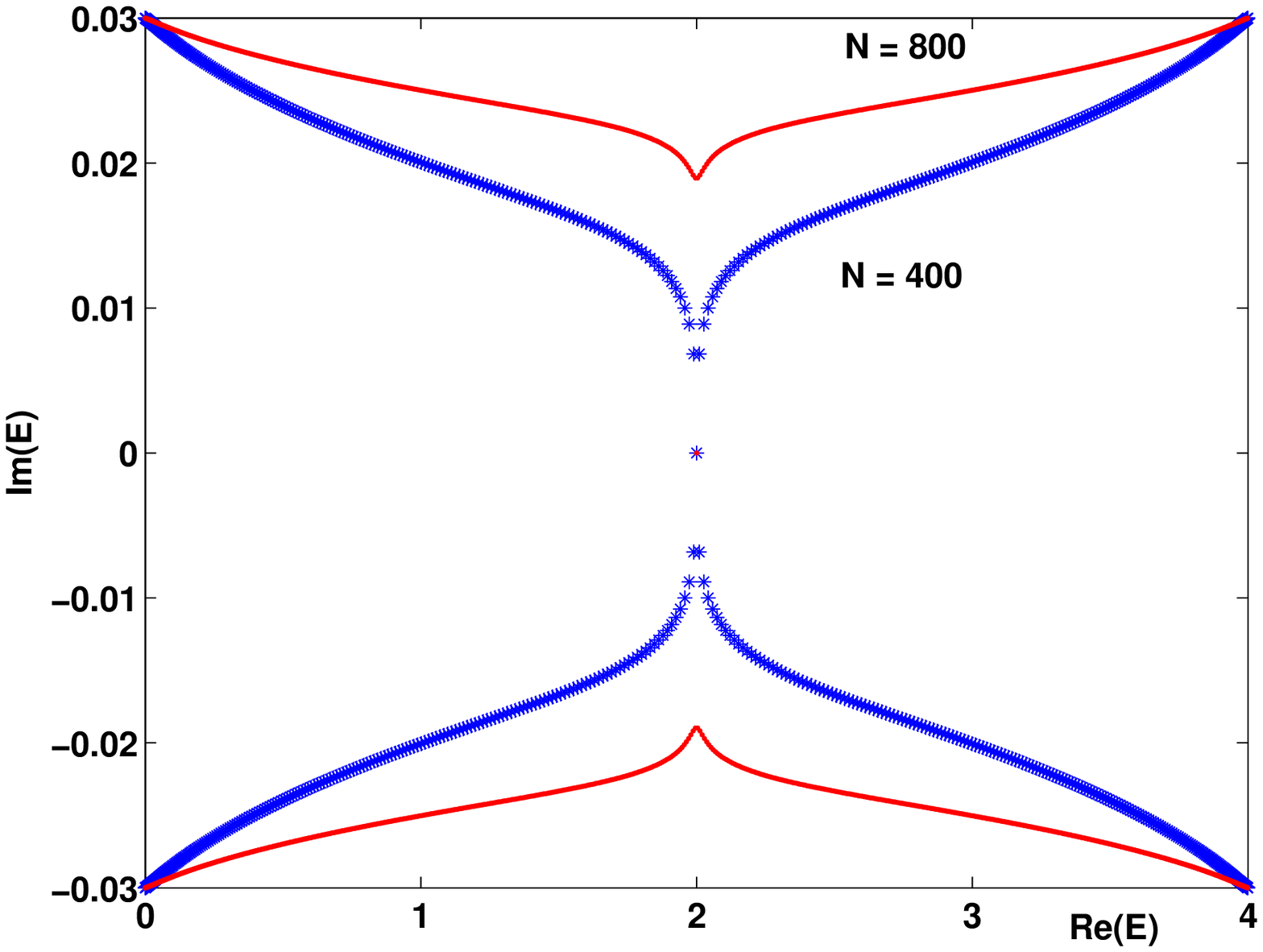}
\includegraphics[width=70mm,height=50mm]{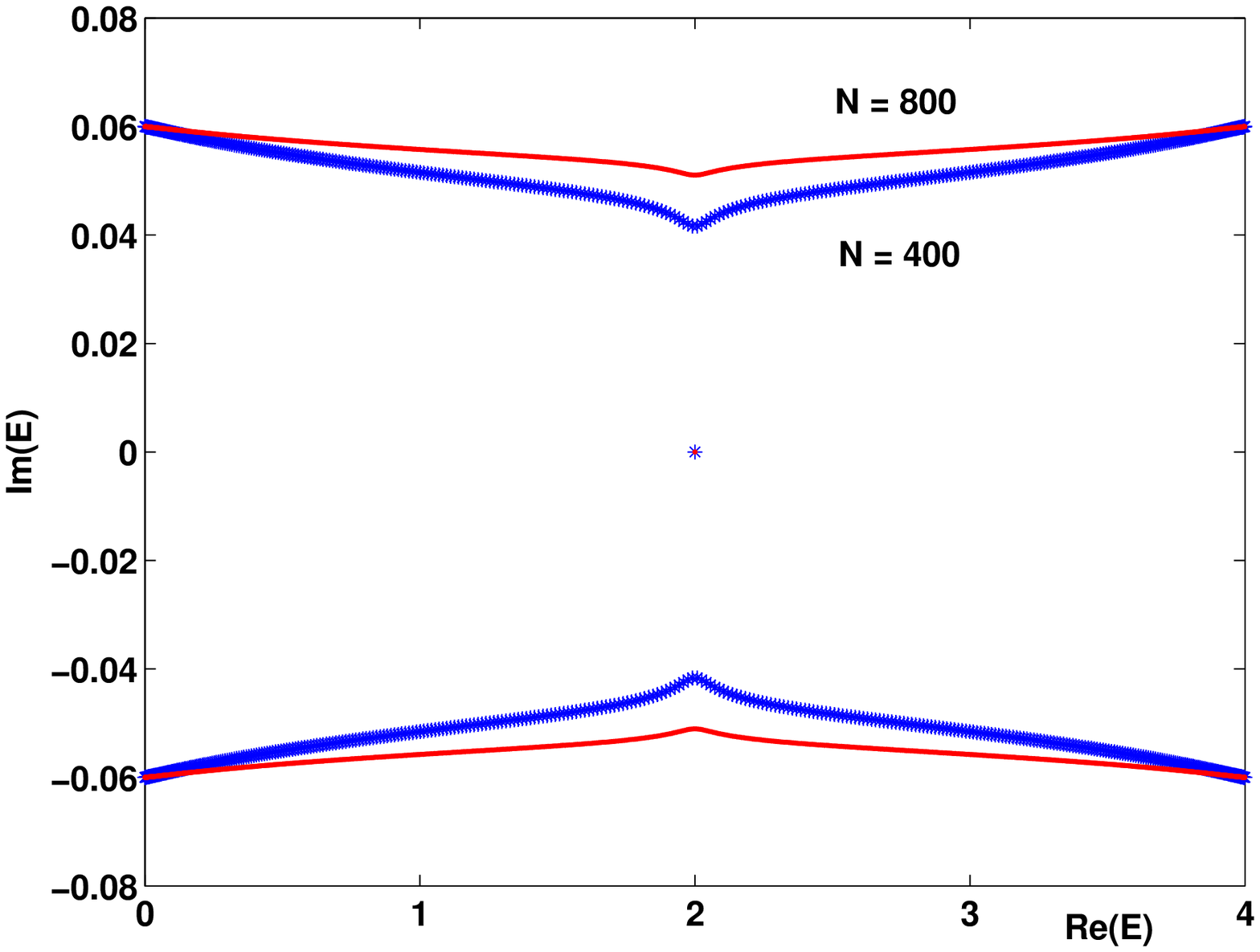}
\end{center}
\caption{(Color online) Eigenvalues of the linear lattice (\ref{eqn7}),
for the cases of $\gamma=0.03$ (top) and $\gamma=0.06$ (bottom).}
\label{fig2}
\end{figure}

Figure~\ref{fig2} shows eigenvalues of the truncated linear lattice (\ref{eqn7})
for $\gamma = 0.03$ (top) and $\gamma = 0.06$ (bottom). At a first glance, it seems that
the numerical eigenvalues do not correspond to the analytical results above (one eigenvalue at
$E = 2$ and the continuous spectrum at ${\rm Im}(E) = \pm \gamma$ for ${\rm Re}(E) \in [0,4]$).
However, this is an artifact of the truncation of the infinite lattice by a finite number
$N$ of lattice sites subject to the Dirichlet end point conditions. When $N$ is increased from
$N = 400$ (stars) to $N = 800$ (dots), the numerical eigenvalues are strongly affected
and move towards the location of spectrum for the infinite lattice (\ref{eqn7}).

Note that the spectrum of the linear lattice (\ref{eqn7}) is very different
from the spectrum of the linear lattice of dimers:
\begin{eqnarray}
E u_n = -\left(u_{n+1} + u_{n-1} -2 u_n\right) + i \gamma (-1)^n u_n.
\label{eqn8}
\end{eqnarray}
As it is well-known \cite{Sukh,KPZ,instability}, the spectrum of the linear lattice
of dimers (\ref{eqn8}) is purely continuous and located at
\begin{eqnarray}
E = 2 \pm \sqrt{4 \sin^2\left(\frac{\theta}{2}\right)-\gamma^2}, \quad \theta \in [-\pi,\pi],
\end{eqnarray}
which still shows instability for any non-zero value of $\gamma$ because
of the Fourier modes with $\theta$ close to $0$.

{\it A potential with a localized $\cP \cT$-symmetric part.}
Lastly, we mention briefly an example in which the $\cP \cT$-symmetric
potential is localized in space, i.e.,
\begin{equation}
\label{eqn9}
V_n = {\rm sech}^2(n/M)  \left(1 + i \gamma \tanh(n/M) \right),
\end{equation}
where parameter $M$ denotes the width of the potential.
If $\gamma = 0$, the linear eigenvalue problem with
the real potential (\ref{eqn9}) admits a continuous
spectrum for $E \in [0,4]$ and a finite number of simple eigenvalues
for $E > 4$ (because $V_n \geq 0$, no eigenvalues appear for
$E < 0$ for $\gamma = 0$.) When $\gamma$ is small but non-zero, all simple
eigenvalues persist in $\cP \cT$-symmetric potentials
by the perturbation theory similar to the one considered
in \cite{R28a}. On the other hand, since the potential $V_n$
decays exponentially as $|n| \to \infty$, the continuous spectrum
is not affected by the relatively compact perturbations and
is determined by the oscillatory non-decaying eigenfunctions
with real $\theta$ and $E(\theta) = 2 - 2\cos(\theta) \in [0,4]$.
Thus, we conclude that the spectrum of the linear eigenvalue
problem with the $\cP \cT$-symmetric  potential (\ref{eqn9}) is real
at least for small $\gamma > 0$.

\begin{figure}[tbp]
\begin{center}
\includegraphics[width=70mm,height=50mm]{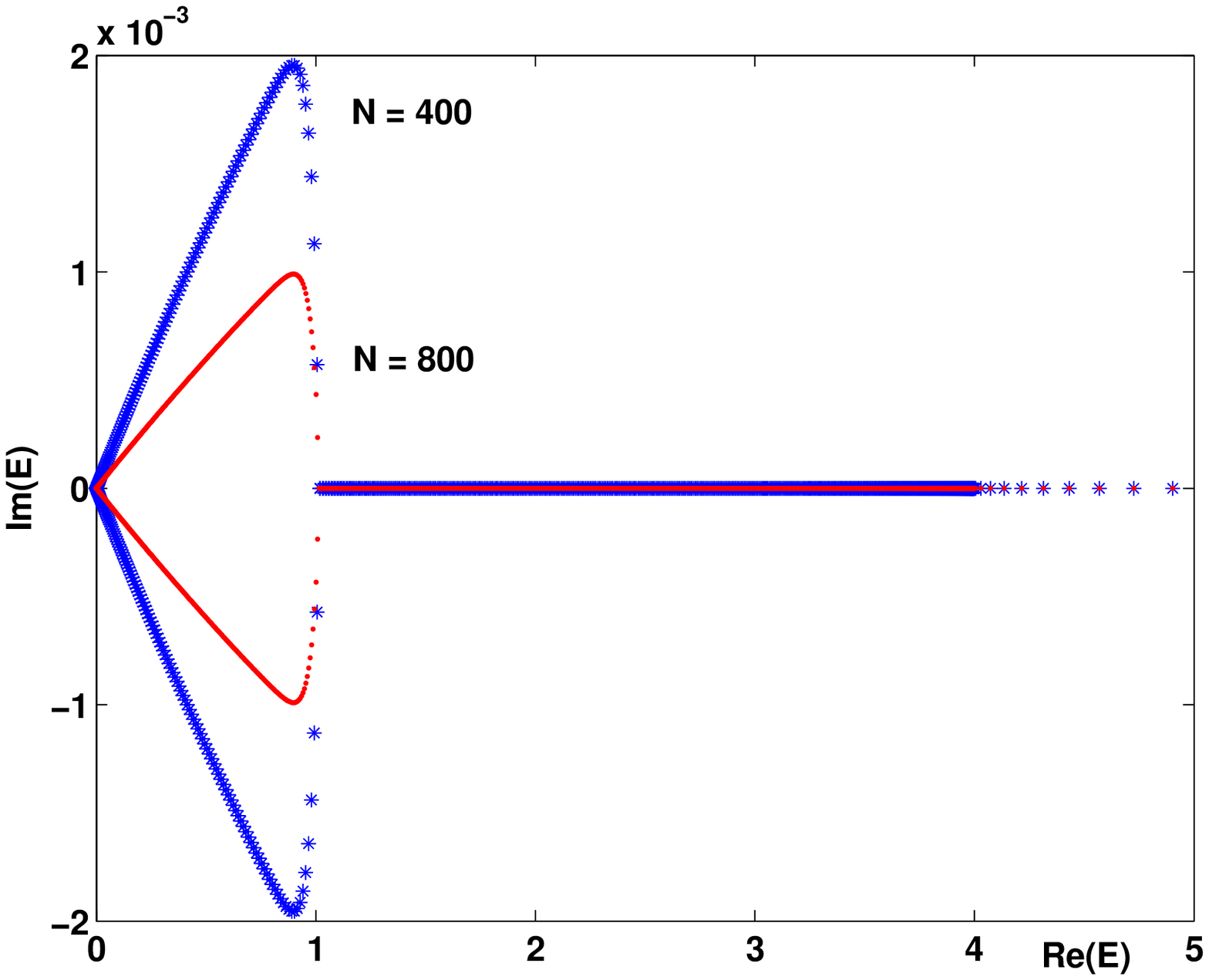}
\includegraphics[width=70mm,height=50mm]{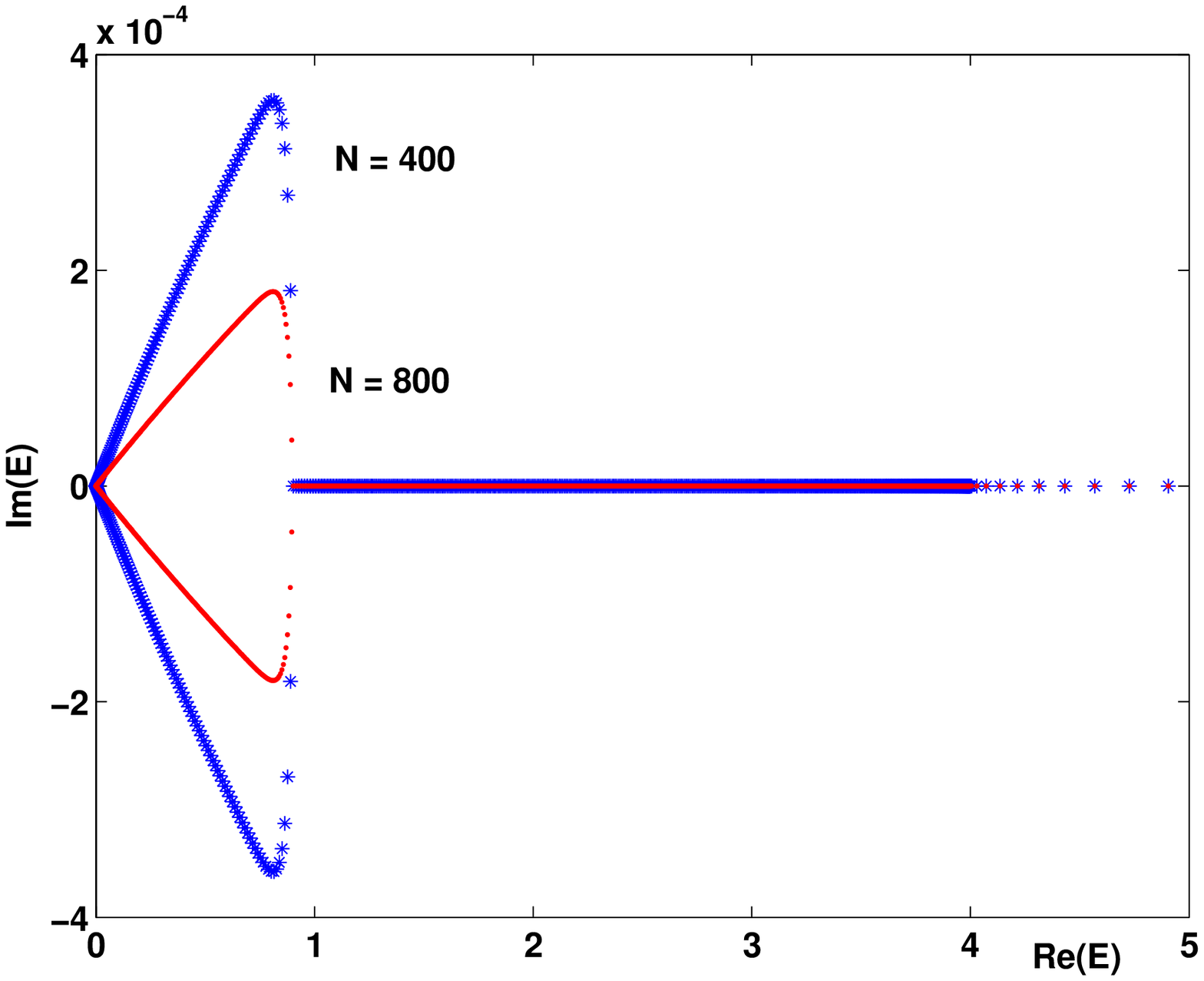}
\end{center}
\caption{(Color online) Eigenvalues of the linear lattice with the potential
(\ref{eqn9}) for $M = 10$ and $\gamma=0.02$ (top) or $\gamma=0.1$ (bottom).}
\label{fig4}
\end{figure}

Figure~\ref{fig4} shows the spectrum of the truncated linear lattice with
the potential (\ref{eqn9}) for $M = 10$ and $\gamma = 0.02$ (top) or
$\gamma = 0.1$ (bottom). Simple eigenvalues for $E > 4$ persist on the real axis
for small values of $\gamma$. Although it seems that the continuous spectrum
becomes complex near $E = 0$, this is again a numerical artifact, as adding
more lattice sites $N$ reduces
the imaginary part of the eigenvalues in the truncated lattice.
As $N$ increases, we anticipate that
the eigenvalues approach the real segment $[0,4]$ for the spectrum of 
the linear lattice with the potential (\ref{eqn9}).

Note that the same conclusion holds also for the compact
$\cP \cT$-symmetric  potentials such as the one corresponding
to the embedded $\cP \cT$-symmetric  defect~\cite{lepri}:
\begin{eqnarray}
V_n = i \gamma (\delta_{n,0} - \delta_{n,1}) u_n,
\label{eqn8a}
\end{eqnarray}
where $\delta_{n,m}$ is the Kronecker delta function.
In this case, the eigenvalue spectrum spans again the interval $[0,4]$
{\it for all} values of $\gamma \in [-\sqrt{2},\sqrt{2}]$ in the
infinite lattice size (i.e., $N \rightarrow \infty$) limit.
As an interesting aside, we mention that the critical
point in this case ($\gamma=\sqrt{2}$) is different from
the critical point (of $\gamma = 1$)
for the $\cP \cT$-phase transition of a single dimer. In any case,
we conclude that the compact
support of the $\cP \cT$-symmetric defect enables a genuinely real
spectrum at least for small values of gain/loss coefficient in the infinite
lattice limit, as was the case for the potential
with the localized $\cP \cT$-symmetric part.


{\it Conclusions.} We have considered the effect of discreteness
on the spectral properties of linear $\cP\cT$-symmetric systems,
characterized by spatially extended gain/loss and have contrasted
these to the continuous $\cP\cT$-symmetric systems. 
We found that discreteness features generic instability of
linear $\cP\cT$-symmetric lattices with extended gain/loss that can be
unveiled by means of Fourier techniques, in conjunction with suitable
analysis of the resulting differential equations in Fourier space.
We have also illustrated some remarkable differences in spectra
between the finite lattice with $N$ sites and the infinite lattice with
$N \rightarrow \infty$.

The considered models with extended gain/loss
featured generic instability properties that emerge essentially as soon
as $\gamma \neq 0$, i.e., their $\cP\cT$-phase transition 
occurs at $\gamma=0$ for all these linear lattices. Notice that the
instability can not be avoided in the respective nonlinear settings,
since $\cP\cT$-phase transition is a purely linear phenomenon, and
nonlinearity can not arrest the exponential growth of the unstable
eigenstates. Nevertheless,
this instability can be dramatically avoided 
if the gain/loss are instead localized or compact: in this case,
a purely real spectrum of the infinite linear lattice can emerge for small
values of gain/loss coefficient $\gamma$.

There are many interesting directions for future studies.
First, providing a {\it sharp} criterion about
the existence of real eigenvalues in infinite lattices depending on
the localization rate of the $\cP\cT$-symmetric potentials
would be an extremely interesting condition both from a mathematical and
physical perspective.
Furthermore, the difference in spectra between finite versus infinite lattices
is another problem that merits additional investigation. It would also be interesting
to extend these considerations to higher dimensional lattices.

\end{document}